# Small Language Models Reshape Higher Education: Courses, Textbooks, and Teaching

Jian Zhang[1]; Jia Shao[2*]

[a] School of Geophysics and Geomatics, China University of Geosciences, Wuhan 430074, China

[b] College of Informatics, Huazhong Agricultural University, Wuhan 430070, China

*Corresponding to Dr. Shao (shaojia@mail.hzau.edu.cn)

**Abstract** While large language models (LLMs) have introduced novel paradigms in science and education, their adoption in higher education is constrained by inherent limitations. These include a tendency to produce inaccuracies and high computational requirements, which compromise the strict demands for accurate and reliable knowledge essential in higher education. Small language models (MiniLMs), by contrast, offer distinct advantages in professional education due to their lightweight nature and precise retrieval capabilities. This research takes "Atmospheric Physics" as an example. We established a specialized corpus and image repository by gathering over 550,000 full-text PDFs from over 130 international well-respected journals in Earth and environmental science. From this collection, we extracted over 100 million high-quality sentence-level corpus and more than 3 million high-resolution academic images. Using MiniLMs, these resources were organized into a high-dimensional vector library for precise retrieval and efficient utilization of extensive educational content. Consequently, we systematically redesigned the courses, textbooks, and teaching strategies for "Atmospheric Physics" based on MiniLMs. The course is designed as a "interdisciplinary-frontier" system, breaking down traditional boundaries between atmospheric science, space science, hydrology, and remote sensing. Teaching materials are transformed from static, lagging text formats into a dynamic digital resource library powered by MiniLM. This library features real-time and expert-verified information. For teaching methods, we have designed a question-based learning pathway centered on "problem identification, resource exploration, and evidence-based argumentation," facilitated by the efficient retrieval of MiniLMs. This paradigm promotes a shift from passive knowledge transfer to active cognitive development. Consequently, this MiniLM-driven "Atmospheric Physics" course not only demonstrates a specific avenue for "AI for education" but also provides both theoretical foundations and practical examples for the collaborative advancement of courses, textbooks, and teaching strategies in higher education.

# 小语言模型重构高等教育课程、教材、教法

张健 [1]；邵佳 [2*]

[1] 地球物理与空间信息学院,中国地质大学, 武汉 430074, 中国

[2] 信息学院,华中农业大学, 武汉 430070, 中国

通讯作者：邵佳（shaojia@mail.hzau.edu.cn）

**摘要** 在人工智能技术迅猛发展的背景下，大语言模型（LLM）虽为科学与教育提供了新范式，但存在幻视及计算成本高昂等固有缺陷，难以满足高等教育领域对知识准确性、前沿性、和可靠性的严苛要求。小语言模型（MiniLM）凭借轻量化及精准检索能力，在专业教育场景中展现独特适配性。本文以《大气物理》为研究载体，通过收集 130 余种地球与环境科学领域国际权威学术期刊的超 55 万份 PDF 全文，解析切割形成超 1 亿条高质量句子级语料及 300 余万张高分辨率学术图片,构建专业化语料库与图像库。依托 MiniLM 将语料库映射为高维向量库，实现海量教育资源的精准定位与高效调用。在此基础上，基于 MiniLM，系统重构《大气物理》的课程、教材与教法。在课程上，通过打通大气、空间、水文、遥感等学科壁垒，构建"跨学科+学科前沿"的课程内容体系。在教材上，以 MiniLM 为驱动，打造"动态更新、多模态交互、可溯源验证"的数字化教材资源库，突破传统纸质教材信息滞后、呈现单一的局限。在教法上，基于 MiniLM 的精准、高效检索，设计"问题导向-资源支撑-探究论证"的探索式教学路径，推动教学从"知识灌输"向"认知建构"转型。该 MiniLM 驱动的《大气物理》课程不仅开辟一条 AI 赋能教育的具体路径，更为高等教育"课程、教材、教法"的协同革新提供理论支撑与实践范式。

**关键词** 小语言模型；课程重构；动态数字教材；问题导向教法

AI 技术与教育教学的深度融合已成为推动高等教育高质量发展的核心引擎。2025 年 1 月，中共中央、国务院印发的《教育强国建设规划纲要（2024—2035 年）》明确提出 "强化学校教育主阵地作用，全面提升课堂教学水平"、"探索数字赋能大规模因材施教、创新性教学的有效途径" 的战略要求；同年 5 月 14 日，教育部部长怀进鹏在世界数字教育大会上进一步强调 "促进智能技术与教育深度融合,构建未来课堂"、"引导学生合理使用人工智能"。在此背景下，如何通过 AI 技术赋能教学变革，破解传统教学中的痛点问题，

成为高等教育领域亟待回应的时代命题。

生成式人工智能（generative artificial intelligence，GenAI）技术的迅猛发展为教育变革提供了全新可能[1]。DeepSeek 等产品不仅推动课堂教学发生深层次变革，提升课堂对话有效性，更能满足学习过程全覆盖的个性化需求，从高效获取已知、深度构建未知两个维度助力教学提质增效。然而，作为 GenAI 核心载体的大语言模型（LLM），在教育领域的规模化应用仍面临显著瓶颈：其一，内容准确性难以保障，模型易出现"一本正经地胡说八道"的现象，生成信息模糊、事实失准的内容，与高等教育领域对知识传递的精准性要求相悖[2][3]；其二，计算成本高昂，大规模参数训练与部署需强大的硬件支撑[4]，难以适配常态化课堂教学场景；其三，过度依赖 LLM 可能削弱师生间的人际联结，导致学生认知出现惰性化、依赖化倾向，违背教育"立德树人"的本质要求。而非生成式小语言模型（MiniLM）为破解上述困境提供了关键路径。与 LLM 相比，MiniLM 具有鲜明的适配优势：首先，其匹配、推荐的内容信息来源明确、可靠性高，无幻视现象；再者，其可运行在现代几乎任意个人电脑上，可在课堂即时高效运行，运算成本显著低于 LLM[5]；最后，其可实现在课堂上，师生共同探究学习，思考前沿。

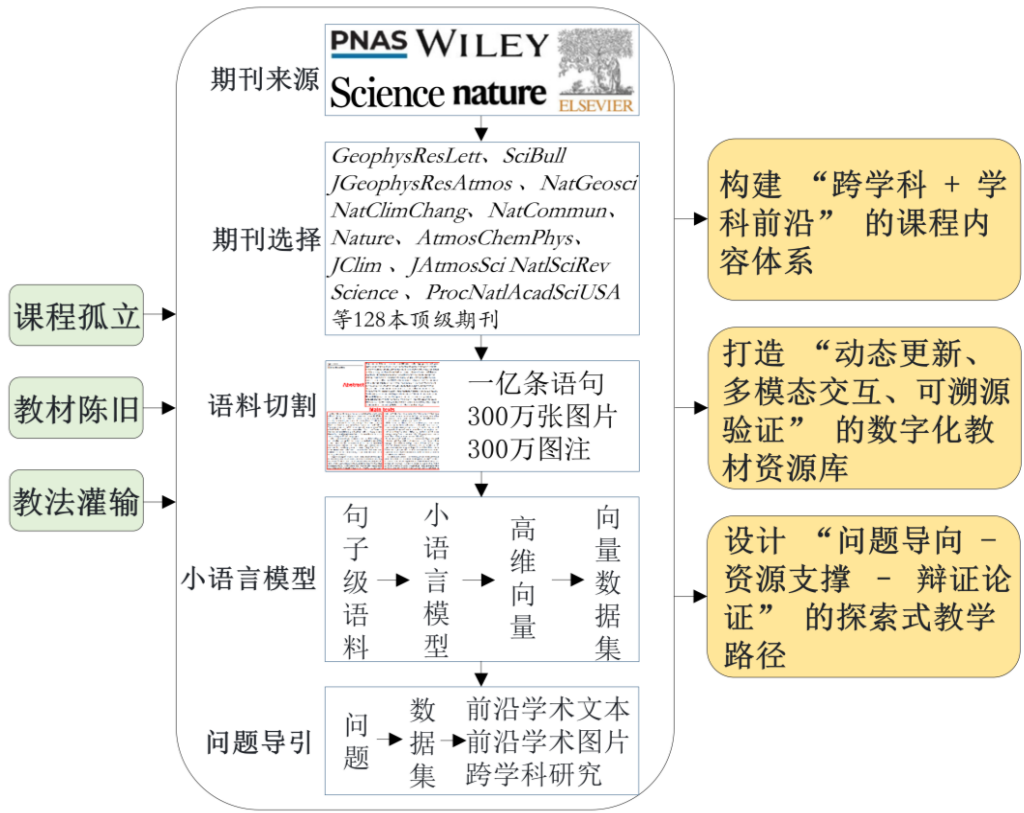

**图 1 小语言模型重构高等教育课程、教材、教法概念图。（左）当前高等专业教育所面临困境。（中）小语言模型耦合科学研究技术路线。（右）实现"语料标准化-检索精准化-应用场景化"高等教育课程、教材、教法智能升级。**

《大气物理》作为高等教育地球科学类专业核心课程，兼具理论性与实践性，其知识体系既涵盖经典物理，又需及时吸纳大气探测、气候变化领域的前沿成果，且与水文、空间、遥感等学科存在紧密的交叉关联。而传统《大气物理》教学面临三大痛点：（1）课程内容学科边界固化，跨学科知识整合不足；（2）

教材更新滞后，难以反映学科前沿；（3）教学方法以灌输式为主，学生探究能力培养薄弱。基于此，本文以《大气物理》课程为研究载体，以"语料标准化-检索精准化-应用场景化"为核心路径，系统设计基于MiniLM驱动的"课程跨学科融合、教材小模型驱动、教法启发式探究"的实践体系，如图1所示。

# 一、课程跨学科融合：构建 "跨学科+学科前沿"的课程内容体系

高等教育课程的系统性与前沿性是保障教学质量的核心，传统《大气物理》课程存在知识点碎片化、学科交叉不足、前沿成果融入滞后等问题，难以适应新时代复合型人才培养需求。非生成式 MiniLM 凭借轻量化及精准检索能力，能深度挖掘学科间的内在关联，为构建 "跨学科+学科前沿" 的课程内容体系提供技术支撑。

## （一）专业化语料库与图像库构建

高质量语料是 MiniLM 发挥效能的前提。本研究聚焦《大气物理》及相关交叉学科，收集了 130 余种地球与环境科学领域国际权威期刊超 55 万份 PDF 全文，涵盖《Nature Geoscience》、《Geophysical Research Letters 》、《Earth and Planetary Science Letters》、《Journal of Climate》《Remote Sensing of Environment》等国际权威出版物，时间跨度覆盖 2000-2025 年，确保语料的权威性与时效性。针对 PDF 格式复杂、信息提取难度大的问题，建立标准化处理流程：在文本处理方面，优先提取摘要与正文内容，剔除参考文献、作者简介等非核心模块，采用模拟人类阅读习惯的页面分割算法处理多列文本，最终切割形成超 1 亿条高质量句子级语料；在图像处理方面，重点提取论文中的实验观测图、原理示意图、数据统计图等学术图片，最终构建包含 300 余万张高分辨率图片的图像库，为"跨学科+学科前沿"课程内容体系构建奠定坚实的数据基础。具体技术细节见[6]。

## （二）跨学科知识体系整合

非生成式 MiniLM 通过语义检索技术，计算人类提问与海量原始语料之间的相关性，能精准捕获《大气物理》与其他学科的知识关联，打破传统课程的学科边界。而大量的 MiniLM，可非常轻易免费下载，例如 *all-MiniLM-L6-v2*、*all-mpnet-base-v2* 等语言模型。具体技术细节见[6]。

例如，通过提问"地震和降水存在很强的相关"，能匹配出大量描述地震和降水相关性的原始语料，精准捕获大气科学与地球物理之间的关联；通过提问"地下水与降水关系非常密切"，能精准将大气与水文链接；通过提问"降水与空间天气事件存在强相关性"，能有效将大气物理与空间科学相耦合。例如，图 2 通过提问"The wind energy is awesome"，则可有效将大气物理与新能源相链接。图 3 则可进一步匹配出风能在所有能源中占比的分布图。

此外，通过无监督聚类算法，对海量语料进行主题分类，能有效构建不同学科之间的关联。比如，当筛选所有包含"precipitation/rain"(降水)的语料后，获得约 180 万句子，通过无监督聚类，则可自动追踪降水在整个地球系统中的角色（降水与生态、降水与地理、降水与环境、降水与空间环境、降水与城市、降水与水文过程等）。更多技术细节见[6]。

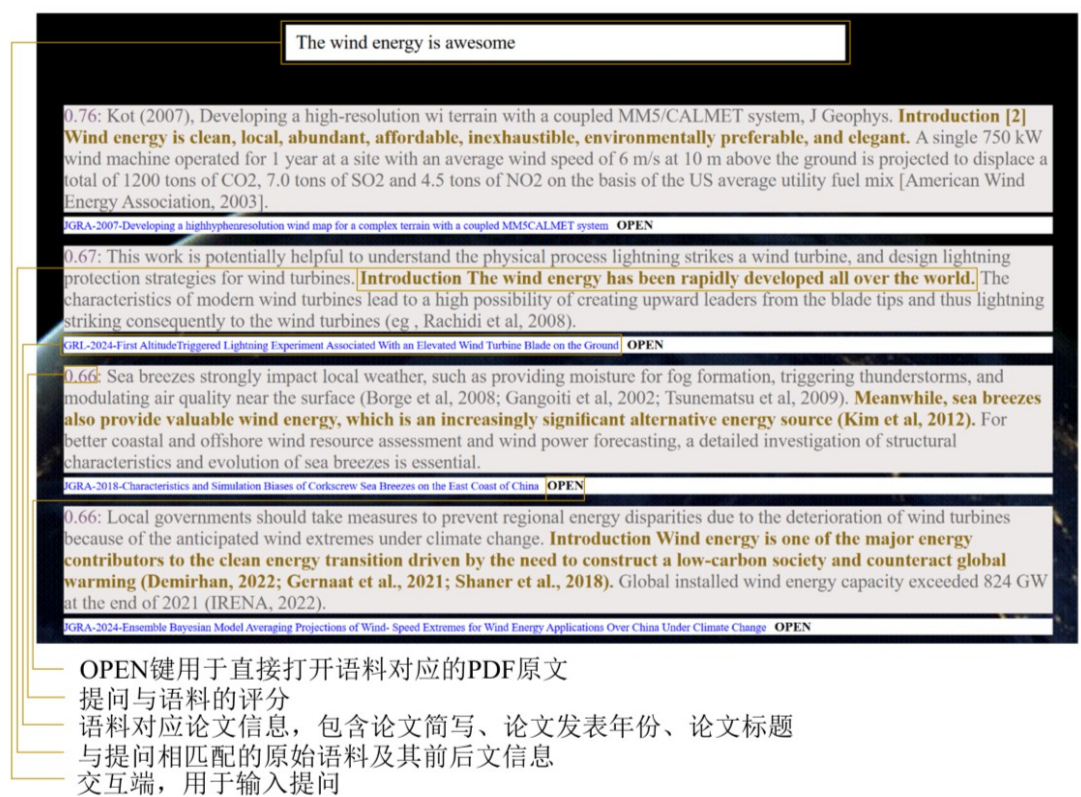

**图 2** MiniLM 与大气物理专业教育耦合的文本检索软件界面。通过一个非常简单的提问("The wind energy is awesome")，可快速匹配出批量高度相关的来源于国际权威顶级期刊的原始语料。语料的前后文、语料来源、语料所对应的 PDF 原文，均可通过软件轻易获知。

**（三）学科前沿成果动态融入**

当下，伴随着学术论文的日益更新与增长，学科知识复杂性、深度等正爆炸式增长。语料库数据系统应自动追踪核心权威论文，做到及时、自动更新。依托语料库系统的自动更新能力，能够将大气物理领域的前沿成果及时纳入课程体系。此后，通过 MiniLM 的检索，可针对性筛选最新研究成果。例如，当提问"地震和降水存在很强的相关"后，MiniLM 可能会推荐从 2000-2025 年，1000 条类似句子，当引导学生只关注最近三年（2023-2025）的研究成果，则可获知最前沿学术动态。以此揭示前沿成果与经典理论的内在联系，帮助学生构建"经典理论-前沿进展-未来方向"的完整知识体系，培养其学术视野与创新思维。

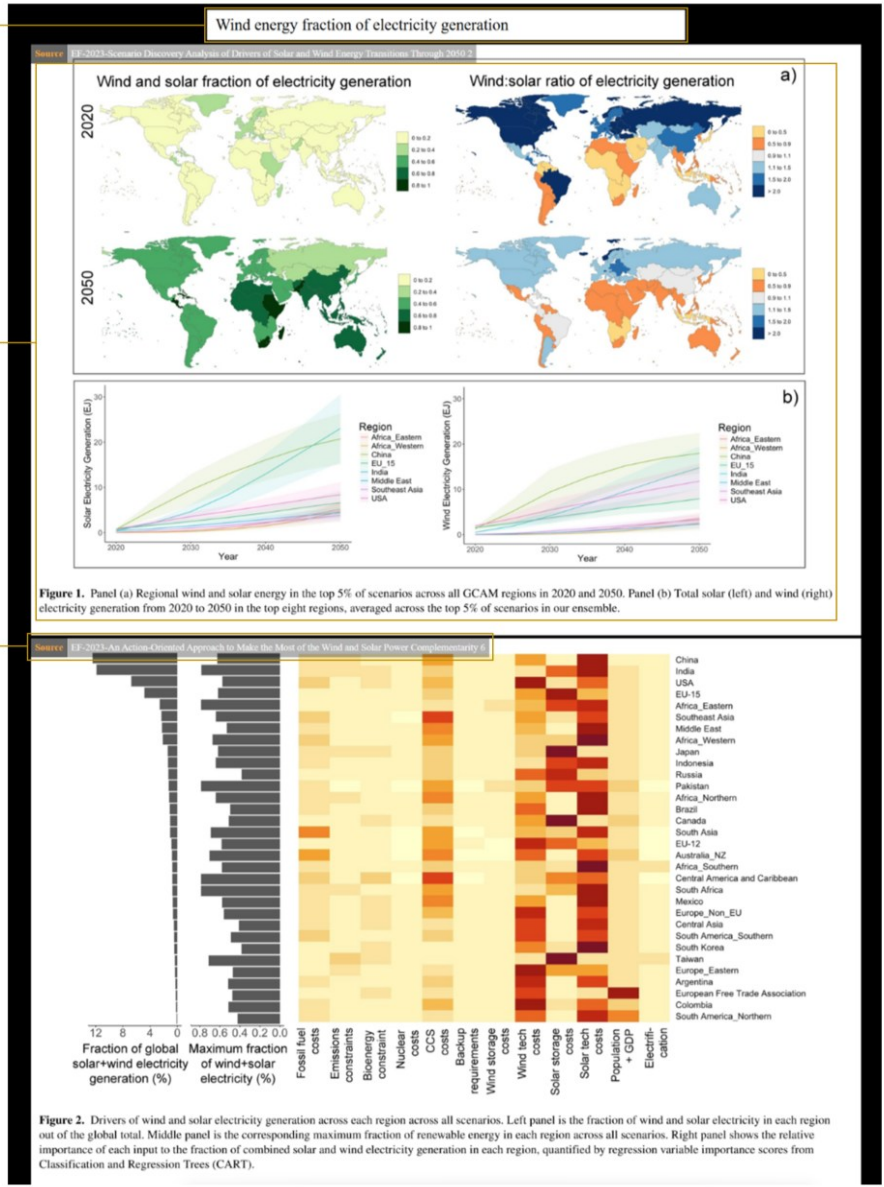

**图 3** MiniLM 与大气物理专业教育耦合的图片检索软件界面。通过一个非常简单的提问("Wind energy fraction of electricity generation"),可快速匹配出批量高度相关的来源于国际权威顶级期刊的原始图片。图片来源在每张图片左上角可轻易获知。

## 二、教材小模型驱动:打造 "动态更新、多模态交互、可溯源验证" 的数字化教材资源库

传统《大气物理》纸质教材存在信息容量有限、呈现形式单一、更新周期漫长等固有缺陷,已无法满

足数字化时代的教学需求。MiniLM 的应用推动教材形态从静态文本向"动态更新、多模态交互、可溯源验证"的数字化资源库转型，为教与学提供更具灵活性与针对性的支撑。

## （一）动态更新机制设计

数字化教材资源库以 MiniLM 为核心驱动，建立了常态化更新机制。模型通过对接权威期刊数据库的 API 接口，实时监测大气科学及相关交叉学科的最新研究成果，自动抓取符合课程要求的学术论文、研究报告等资源。经"机器筛选-专家审核"的双重把关流程，将优质内容整合至教材资源库中，实现教材内容的动态迭代。例如，当学界在大气边界层物理研究中取得新的观测突破时，模型可快速提取核心结论、数据图表等关键信息，更新至"大气边界层"章节，并标注更新时间与来源文献，解决传统教材更新滞后的问题。同时，资源库支持用户自定义更新频率与主题范围，教师可根据教学进度与学生需求，灵活调整教材内容的更新方向。

## （二）多模态交互体系构建

依托 MiniLM 的强大匹配能力，数字化教材资源库整合了文本、图像、表格等多种资源形态，打造沉浸式学习体验。在文本资源方面，提供不同深度的知识解读，包括基础概念阐释、重难点解析、学术论文节选等，满足不同层次的学习需求；在图像资源方面，通过模型的图像语义检索技术，将 300 余万张高分辨率学术图片与知识点精准关联，例如讲解"云系分类"时，同步呈现卫星云图、微观云滴图像、云系演化示意图等多类型图片，帮助学生直观理解抽象概念。

## （三）可溯源验证功能实现

为保障知识传递的准确性与严谨性，数字化教材资源库构建了完善的可溯源验证机制。MiniLM 在整合各类资源时，严格保留原始文献信息，包括期刊名称、发表年份、链接等，所有知识点均标注明确的来源依据。例如，在呈现"广州市 2017 年最大日降水量为 524.1mm"这一数据时，同步标注数据来源于《Geophysical Research Letters》的相关学术论文，学生可通过链接直接查阅原文，验证数据的真实性与可靠性。对于争议性学术问题，模型会整合不同观点的研究成果，并清晰标注各观点的支撑证据与适用场景，引导学生进行批判性思考。这种可溯源验证机制不仅契合学术研究的严谨性要求，更能培养学生的学术规范意识与批判性思维能力。

# 三、探索式教法：设计 "问题导向-资源支撑-探究论证"的探索式教学路径

传统《大气物理》教学多采用"教师讲授-学生接受"的灌输式教法，难以激发学生的主动思考与创新

思维。MiniLM 的应用重构了教学互动模式,通过设计"问题导向-资源支撑-探究论证"的探索式教学路径,推动教学从"知识灌输"向"认知建构"转型,强化学生的科学探究能力与创新思维。

问题设计是探索式教学的起点。教师基于《大气物理》课程核心知识点、学科前沿热点及实际应用需求,结合学生认知水平,设计具有启发性、挑战性的探究问题。问题设置遵循"由浅入深、由表及里"的原则,既包括基础原理应用类问题(如"大气静力方程在不同海拔高度的适用性"),也涵盖跨学科综合类问题(如"遥感技术如何助力大气污染溯源?"),还包含前沿探索类问题(如"全球变暖背景下,中纬度地区大气环流异常的物理机制")。

例如,图 4 试图寻找风速随着气候变化在减弱的证据。我们能很轻易引导学生获取大量相关研究成果,且均发表于最近五年。而图 5 则展示了一个完全相反的例子,试图引导学生,去寻找风速随着气候变化在增强的证据。事实是,学术界同时存在两种可能完全相反的声音,这说明该领域尚未达成共识。通过这种方式,能很好培养学生的辩证思想,也增强学生参与学术前沿的信心。

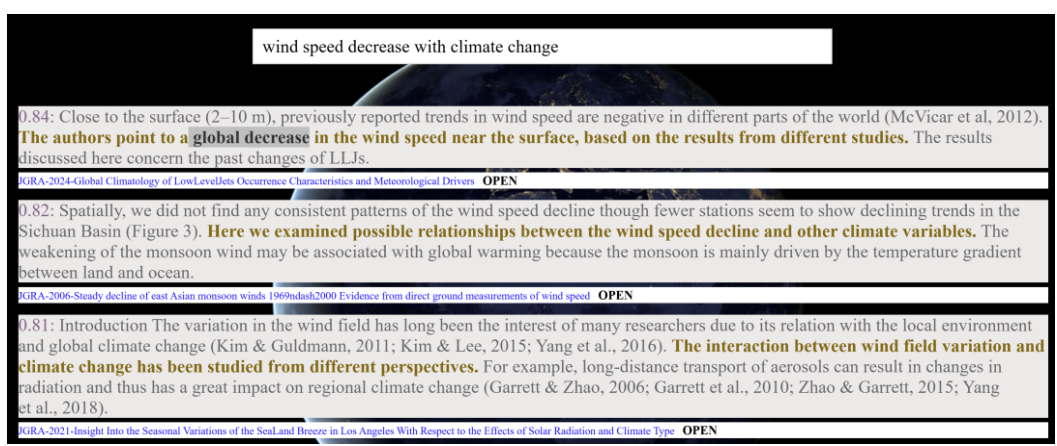

**图 4** 与图 1 类似,当提问为"wind speed decrease with climate change",则可快速筛选出大量论证风速随着气候变化在减弱的相关学术研究成果。

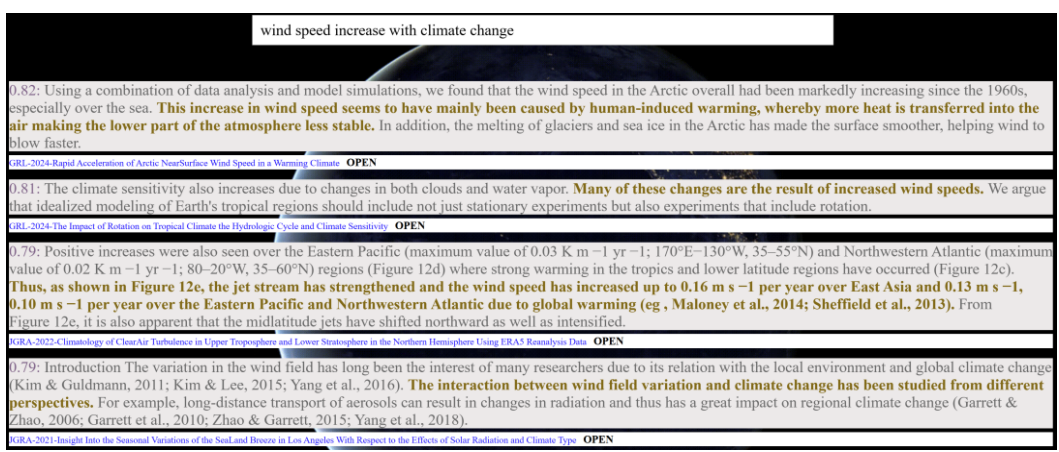

**图 5** 与图 4 类似,而当提问为"wind speed increase with climate change"时,则可快速筛选出大量论证风速随着气候变化在增强的相关学术研究成果。

# 总结

本研究以《大气物理》课程为载体，系统探究了 MiniLM 在"课程、教材、教法"重构中的应用机制与实践路径，构建了"课程跨学科融合、教材小模型驱动、探索式教法"的实践体系，取得了显著成效。

研究表明，MiniLM 凭借轻量化特性与精准检索能力，有效弥补了 LLM 在教育应用中的固有缺陷，为高等教育专业课程的智能化升级提供了可行方案。在课程层面，通过专业级语料库与图像库支撑，实现了跨学科整合与前沿成果动态融入，构建了结构化的课程内容体系；在教材层面，打造了动态更新、多模态交互、可溯源验证的数字化教材资源库，突破了传统纸质教材的局限；在教法层面，设计了"问题导向-资源支撑-探究论证"的探索式路径，推动教学从"知识灌输"向"认知建构"转型。

研究的核心创新在于确立了"语料标准化-检索精准化-应用场景化"的核心路径：通过权威学术资源的标准化处理，保障了模型应用的基础质量；借助语义检索、无监督聚类等核心技术，实现了教育资源的精准匹配与高效调用；通过课程、教材、教法的场景化应用，将技术优势转化为教学实效。这一路径不仅为《大气物理》课程的智能化升级提供了具体方案，更为高等教育同类专业课程的协同革新提供了可复制、可推广的实践范式。相关资料下载链接:: https://pan.baidu.com/s/1rSTbJHHdvCuZhSXJ0pN-dQ（提取码 y5zh）。

# 参考文献